\documentclass[showpacs,aps,prd,nofootinbib,floatfix,amsmath,amssymb]{revtex4}
\usepackage{graphicx}
\usepackage[utf8x]{inputenc}
\usepackage{color}
\usepackage{subfig}
\usepackage{multirow}
\usepackage{slashed}

\begin{document}

\title{Lepton masses and mixing in a scotogenic model}


\author{E. C. F. S. Fortes}%
\email{elaine@ift.unesp.br}
\affiliation{
	Universidade Federal do Pampa\\
	Rua Luiz Joaquim de Sá Brito, s/n, Promorar,\\ Itaqui - RS, 97650-000,
	Brazil
}

\author{A. C. B. Machado}%
\email{ana@ift.unesp.br}
\affiliation{
Laboratorio de F\'{\i}sica Te\'{o}rica e Computacional\\
Universidade Cruzeiro do Sul -- Rua Galv\~{a}o Bueno 868 \\
S\~{a}o Paulo, SP, Brazil, 01506-000
}

\author{J. Monta\~no}%
\email{montano@ift.unesp.br}
\affiliation{
	Instituto  de F\'\i sica Te\'orica--Universidade Estadual Paulista \\
	R. Dr. Bento Teobaldo Ferraz 271, Barra Funda\\ S\~ao Paulo - SP, 01140-070,
	Brazil
}

\author{V. Pleitez}%
\email{vicente@ift.unesp.br}
\affiliation{
Instituto  de F\'\i sica Te\'orica--Universidade Estadual Paulista \\
R. Dr. Bento Teobaldo Ferraz 271, Barra Funda\\ S\~ao Paulo - SP, 01140-070,
Brazil
}

\date{05/26/17}
%


\begin{abstract}
We consider an extension of the standard model with three Higgs doublet model and $S_3\times \mathbb{Z}_2$ discrete symmetries. Two of the scalar doublets are inert due to the $\mathbb{Z}_2$ symmetry. We have calculated all the mass spectra in the scalar and lepton sectors and accommodated the leptonic mixing matrix as well. We also show that the model has scalar and pseudoscalar candidates to dark matter. Constraints on the parameters of the model coming from the decay $\mu\to e\gamma$ were considered and we found signals between the current and the upcoming experimental limits, and from that decay we can predict  the one-loop $\mu\to ee\bar{e}$ channel.

\end{abstract}

\pacs{
14.60.Pq, 
14.60.St, 
13.35.Bv 
}

\maketitle

\section{Introduction}
\label{sec:intro}

Although since 2012 we know that there exist a neutral spin-0 resonance with properties (mass and couplings) that are compatible, within the experimental error, with the Higgs boson of the standard model (SM)~\cite{Aad:2012tfa,Chatrchyan:2012xdj}, the data do not exclude the existence of more scalar fields and almost all extensions of the SM include extra Higgs doublets. This is the reason for considering multi-Higgs models. Moreover, although many scalar doublets may exist in nature, it is possible that only one of them is the responsible for the electroweak spontaneous symmetry breaking and the generation of the charged fermion masses.
In this case, the other scalar multiplets may be inert ones: they do not couple to fermions, do not contribute to the vector bosons masses, and interact only with vector bosons and with other scalars. This possibility was put forward many years ago in Ref.~\cite{Deshpande:1977rw} in which a $\mathbb{Z}_2$ symmetry was imposed to keep inert one of the doublets.

On the other hand, an indication that there must be physics beyond the SM is the origin of the neutrino masses. In fact, the generation of masses smaller than 0.1 eV demand the introduction of new degrees of freedom even in the context of the gauge symmetries of the SM. An interesting possibility is that the extra scalar fields that may exist as an extension of the SM also induce the appropriate neutrino mass. In particular,
neutrino mass generation in a model with two doublets, being one of them inert, was considered by Ma~\cite{Ma:2006km}. This is the so called scotogenic mechanism for generating neutrino masses through one-loop corrections involving the inert neutral components.

One interesting feature of the mechanism is that it includes by construction one or more dark matter (DM) candidates, or the implementation of the baryon asymmetry in the Universe, relating in this way three of the more important problems in elementary particle physics: the generation of the neutrino masses, the nature of the DM, and the observed asymmetry between matter and anti-matter, see Ref.~\cite{Aoki:2014lha} and references therein. Moreover, the existence of many components of DM is interesting by their own. In this case DM may decay from heavier to lighter components, and also co-annihilate in two dark particles~\cite{Dienes:2014via}. In fact, this might be the only possibility to accommodate several astrophysical observations. For instance, the electron/positron excesses observed by many experiments and recently confirmed by AMS-02~\cite{Accardo:2014lma} and the excess of gamma rays peaking at energies of several GeV  from the region surrounding the Galactic Center~\cite{Goodenough:2009gk,Daylan:2014rsa} must need at least two component DM. Moreover, the latter case can avoid
some constraints on the one component DM from the AMS-02 data~\cite{Geng:2014dea}. Notwithstanding there are alternative interpretations for these gamma rays excess, see \cite{Abazajian:2015raa,Fermi-LAT:2016uux}. It is interesting that the one-doublet inert model has at least two DM components. However see \cite{Liu:2017drf}.

Here we will work out a similar mechanism but in the context of the model with two inert scalar doublets proposed in Ref.~\cite{Machado:2012ed}.
Moreover, the inert character is due to the $\mathbb{Z}_2$ symmetry, and the $S_3$ symmetry makes the scalar potential more predictive and easier to be analyzed. Although three right-handed neutrinos are introduced, the active neutrino masses do not arise through the type-I seesaw mechanism but via the scotogenic mechanism, at the 1-loop level. We need to add two real singlet scalar fields in order to accommodate the charged lepton masses and the Pontecorvo-Maki-Nakagawa-Sakata (PMNS) mixing matrix. In the context of inert doublet models, the latter issue is to the best of our knowledge done for the first time.

The outline of this paper is as follows. In the next section we discuss the model while in Sec.~\ref{sec:scalars} analyse the scalar sector. Lepton mass matrices and the leptonic mixing matrix is shown in Sec.~\ref{sec:masses}. In Sec.~\ref{sec:dm} we show that the model provides a multi-component DM spectra, although we do not consider the most general case.
In Sec.~\ref{sec:decays} we consider the decays $\mu\to e\gamma$, Subsec.~\ref{subsec:muegamma}, and $\mu\to e\bar{e}e$ in Subsec.~\ref{subsec:meee}.
 Our conclusions appear in Sec.~\ref{sec:con}.

\section{The Model}
\label{sec:model}

In Ref.~\cite{Machado:2012ed} it was proposed an extension of the electroweak SM with three Higgs scalars $H_{1,2,3}$ transforming as doublets under $SU(2)$ and having $Y=+1$.  One $H_1\equiv S$ transforms as singlet of $S_3$, and the others as  doublets, $D=(H_2,H_3)~\equiv~(D_1,D_2)$.
Here we will extend the model of Ref.~\cite{Machado:2012ed} by adding three right-handed sterile neutrinos, $N_{1R}$ transforming as singlet,  and $N_d=(N_{2R},N_{3R})$ transforming as doublets of $S_3$, and two real scalar singlets of $SU(2)$ ($Y=0$) but doublets of $S_3$, $\zeta_d=(\zeta_1,\zeta_2)$. See the other quantum numbers in Table~\ref{table1}.
The vacuum alignment is given by $\langle S \rangle=v_{SM}/\sqrt{2}$, and  $\langle D_1,D_2\rangle=0$, $\langle\zeta_{1,2}\rangle=v_\zeta$. In the charged lepton and quark sectors, all usual fields of SM transform as singlet under $S_3$.

With these fields the Yukawa interactions in the lepton sector, invariant under the gauge, $S_3$ and $\mathbb{Z}_2$ symmetries (see Table~\ref{table1}) are given by
\begin{eqnarray}\label{yukawa1}
-\mathcal{L}^{leptons}_{Yukawa}&=& G^l_{ij}\bar{L}_{i}l_{jR}S+G^\nu_{id}\bar{L}_{i} [ N_d\, \tilde{D}]_1+\frac{1}{\Lambda} G^\nu_{is}\bar{L}_{i} [N_s[\tilde{D} \zeta_d]_{1^\prime}]_1    \\ \nonumber&+& \frac{1}{2} M_s \overline{N^c_s} N_s + \frac{1}{2} M_d [\overline{N^c_d} N_d]_1 +H.c.,
\end{eqnarray}
where $i,j=e,\mu,\tau$ (we omit summation symbols), $L_i(l_{iR})$ and $N_{s,d}$ denote the usual left-handed lepton doublets (right-handed singlets) and the right-handed neutrinos, respectively;  $[\tilde{D}\,\zeta_d]_{1^\prime}=\tilde{D}_1\zeta_2-\tilde{D}_2\zeta_1$. $[N_d\,D]_1=N_{2R}D_1+N_{3R}D_2$, according to the $S_3$ multiplication rules, and $\tilde{D}_{1,2}=i\tau^2D^*_{1,2}$. Notice that the doublets $D$ and $\zeta_d$ couple only with neutrinos. We assume that $\langle\zeta_{1,2}\rangle\lesssim \Lambda$, where $\Lambda$ is an energy scale much larger than the electroweak one. It is also interesting to note that the right-handed neutrinos in the $S_3$ doublet are mass degenerated, with mass $M_d$, which is different from the mass of the right-handed neutrino in singlet of $S_3$, which has a mass denoted by $M_s$.
Notice that at three level active neutrinos are still massless.

\begin{table}[ht]
\begin{tabular}{|c|c|c|c|c|c|c|c|}\hline
Symmetry & $L_{i}$ &  $l_{jR}$ & $N_s$ & $N_d$ & $S$ & $D$  & $\zeta_d$  \\ \hline
$S_{3}$  & 1 &  1  & $1^\prime$ & 2 & 1
& 2  & 2  \\ \hline
$\mathbb{Z}_2$ & 1 & 1 & -1 & 1 & 1 & 1 & -1 \\ \hline
\end{tabular}
\caption{Transformation properties of the fermion and scalar fields under $S_{3}$ and $\mathbb{Z}_2$ symmetries. Quarks and charged leptons are singlets of $S_3$ and even under $\mathbb{Z}_2$.}
\label{table1}
\end{table}

\section{The scalar sector}
\label{sec:scalars}

The scalar sector of the model is presented as follows:
\begin{equation}
S=\left (\begin{array}{c}
S^+ \\
\frac{1}{\sqrt2}(v_{SM}+\textrm{Re}S^0+i\textrm{Im}S^0)
\end{array}\right),\quad D_{1,2}=\left( \begin{array}{c}
D^+_{1,2} \\
\frac{1}{\sqrt2}(\textrm{Re}D^0_{1,2}+i\textrm{Im}D^0_{1,2}).
\end{array}\right),
\label{notation}
\end{equation}
plus the singlets $\zeta_i=v_i+\textrm{Re}\zeta_i+i\textrm{Im}\zeta_i$, $i=1,2$.

The scalar potential invariant under the gauge and $S_3\otimes \mathbb{Z}_2$ symmetries is
\begin{eqnarray}
V_{S_3}  &=& \mu^2_sS^\dagger S+\mu^2_d [D^\dagger\otimes  D]_1+ \mu^2_{\zeta_D} [\zeta_D\otimes  \zeta_D]_1 +\lambda_1
([D^\dagger\otimes  D]_1)^2+\mu^2_{12}\zeta_1\zeta_2
+  a_2 [[D^\dagger\otimes D]_{1^\prime}[D^\dagger\otimes
D]_{1^\prime}]\nonumber\\
    &&+a_3[(D^\dagger \otimes D)_{2^\prime}(D^\dagger\otimes D)_{2^\prime}]_1
+a_4(S^\dagger S)^2+
a_5[D^\dagger\otimes D]_1 S^\dagger  S +a_6 [[S ^\dagger D]_{2^\prime} [S^\dagger   D]_{2^\prime}]_1
\nonumber \\
&&+ H.c.]+
a_7 S^\dagger [ D \otimes D^\dagger]_1 S+
b_1 S^\dagger S [ \zeta_D \otimes \zeta_D]_1 + b_2  [D^\dagger\otimes  D]_1 [ \zeta_D \otimes \zeta_D]_1
\nonumber\\
&&+ b_3 [[D^\dagger\otimes  D]_{2^\prime} [ \zeta_D \otimes \zeta_D]_{2^\prime}]_{1}
+  c_1([\zeta_D\otimes  \zeta_D]_1)^2+
c_2 [[\zeta_D\otimes \zeta_D]_{2^\prime}[\zeta_D\otimes
\zeta_D]_{2^\prime}]_1,
\label{potential1}
\end{eqnarray}
with $\mu^2_d>0$ that is guaranteed by the $\mathbb{Z}_2$ symmetry.

We can write Eq.~(\ref{potential1}) explicitly as
\begin{equation}
V(S,D,\zeta_d)=V^{(2)}+V^{(4a)}+V^{(4b)}+V^{(4c)},
\label{potential2}
\end{equation}
where
\begin{eqnarray}
V^{(2)}&=&\mu^2_{SM}S^\dagger S+\mu^2_d (D^\dagger_1D_1+D^\dagger_2 D_2)+\mu^2_\zeta(\zeta^2_1+\zeta^2_2 )+\mu^2_{12}\zeta_1\zeta_2,
\nonumber\\
V^{(4a)} &=&  a_1
(D^\dagger_1D_1+D^\dagger_2 D_2)^2
+  a_2 (D^\dagger_1D_2-D^\dagger_2D_1)^2
\nonumber\\
&&+ a_3[(D^\dagger_1D_2+D^\dagger_2D_1)^2+(D^\dagger_1D_1-D^\dagger_2D_2)^2]
+a_4(S^\dagger S)^2+
a_5  (D^\dagger_1D_1+D^\dagger_2 D_2)S^\dagger  S
\nonumber\\
&&+ a_6[(S^\dagger D_1S^\dagger D_1+S^\dagger D_2S^\dagger D_2)+H.c.]
+
a_7 S^\dagger (D_1D^\dagger _1+D_2 D^\dagger_2) S,
\nonumber\\
V^{(4b)}&=& b_1S^\dagger S(\zeta^2_1+\zeta^2_2)+b_2(D^\dagger_1D_1+D^\dagger_2D_2)(\zeta^2_1+\zeta^2_2)+
b_3[(D^\dagger_1D_2+D^\dagger_2D_1)(\zeta_1\zeta_2+\zeta_1\zeta_2)
\nonumber \\
&&\times(D^\dagger_1D_1-D^\dagger_2D_2)(\zeta^2_1-\zeta^2_2)+H.c.],
\nonumber\\
V^{(4c)}&=&c_1(\zeta^2_1+\zeta^2_2)^2
+ c_2[(\zeta_1\zeta_2+\zeta_2\zeta_1)^2+(\zeta^2_1-
\zeta^2_2)^2],
\label{potential3}
\end{eqnarray}
where we have used $[\zeta_d\,\zeta_d]_{2^\prime}=(\zeta_1\zeta_2+\zeta_2\zeta_1,\zeta_1\zeta_1-\zeta_2\zeta_2)$. Notice that the term $\mu^2_{12}$ breaks softly the $S_3$ symmetry but not the $\mathbb{Z}_2$. Notice also that the $\mathbb{Z}_2$ symmetry forbids trilinear terms in the scalar potential $[D^\dagger\otimes  D]_1\zeta_i$ and $\zeta^3_i$.

From derivation of Eq.~(\ref{potential3}), we obtain the constraint equations:
\begin{eqnarray}
v_{SM}[2\mu^2_{SM}+2a_4v^2_{SM}+b_1(v^2_1+v^2_2) ]&=&0,\nonumber \\
v_1[4\mu^2_\zeta +2b_1v^2_{SM}+4c(v^2_1+v^2_2)
+4\mu^2_{12}\frac{v_2}{v_1}]&=&0,\nonumber\\
v_2[4\mu^2_\zeta+2b_1v^2_{SM}+4c(v^2_1+v^2_2)
+4\mu^2_{12}\frac{v_1}{v_2}]&=&0,
\label{ce1}
\end{eqnarray}
where we have defined $c=c_1+c_2$.
Notice from (\ref{ce1}) that neither $v_1=0$ nor $v_2=0$ are allowed if $\mu^2_{12}\not=0$, hence we have that $v_1\not=0,v_2\not=0$, or $v_1=v_2\equiv v_\zeta$. We have chosen the latter case, so the constraint equations become
\begin{equation}
v_{SM}[\mu^2_{SM}+a_4v^2_{SM}+b_1v^2_\zeta  ]=0\ ,\
v_\zeta[2\mu^2_\zeta+b_1v^2_{SM}+4cv_\zeta  +2\mu^2_{12}]=0.
\label{ce2}
\end{equation}

The scalar potential has to be bounded from below to ensure its stability. In the SM it is easy to ensure the stability of this potential,  we just have to ensure that $\lambda > 0$. In theories in which the number of scalars is increased,  it is more difficult to ensure that the potential is bounded from below, in all directions. A scalar potential has a quadratic form in the quadratic couplings, i.e. $A_{ab}\phi^2_a \phi^2_b $, where $\phi^2_a$ and $\phi^2_b$ represents the scalar fields, $S$, $D$ and $\zeta_d$. If the matrix $A_{ab}$ is copositive it is possible to ensure that the potential has a global minimum.  Assuming a quadratic form (e.g. considering only the quartic terms of the potential) is valid, even if there exist trilinear terms, because in the case where the fields assume large values, the terms of order 2 and 3 are negligible compared to the terms of order 4. For more detail see Refs.~\cite{ping,Kannike:2012pe}. We consider all quartic couplings positive, i.e. all $A_{ij}$ are positive. Below we will denote $B=a_1-2a_2-a_3$ and $C=a_5-2a_6+a_7$. And finally the following limits guarantee that the scalar potential is bounded the from below:
\begin{eqnarray}
&& a_4\geq 0, \;\frac{a_5}{2}\geq 0,\; C \geq 0,\; b_1\geq 0,\; a_1+a_3\geq 0, \nonumber \\&&
b_1+b_2\geq 0,\; B \geq 0,\; c_1+c_2\geq 0.
\label{case1}
\end{eqnarray}

Next, we consider the scalar mass spectra. In the $C\!P$-even sector the mass matrix becomes in block diagonal form with one $3\times3$, $\mathcal{M}_{1R}$ sub-matrix and one $2\times2$ matrix, $\mathcal{M}_{2R}$. The first one in the basis $(\textrm{Re}S^0, \zeta_1,\zeta_2)$ is given by
\begin{equation}
\mathcal{M}^2_{1R}=\left(
\begin{array}{ccc}
2a_4v^2_s & b_1v_{SM}v_\zeta & b_1v_{SM}v_\zeta\\
& -\mu^2_{12}+2cv^2_\zeta & \mu^2_{12}+2cv^2_\zeta  \\
& & -\mu^2_{12}+2cv^2_\zeta
\end{array}\right).
\label{mr1}
\end{equation}
The respective eigenvalues are
\begin{eqnarray}
&& m^2_1= -2\mu^2_{12}, \nonumber \\ &&
m^2_2= \frac{1}{2}\left[a_4v_{SM}^2+2cv^2_\zeta -\sqrt{a_4^2v^4_{SM} + 2 (b_1^2 - 2 a_4 c)v^2_\zeta v^2_{SM} + 4 c^2 v_\zeta^2}\,\right] ,\nonumber \\ &&
m^2_3=\frac{1}{2}\left[[4a_4v_{SM}^2+8cv^2_\zeta]^2 +\sqrt{a_4^2v^4_{SM} + 2 (b_1^2 - 2 a_4 c)v^2_\zeta v^2_{SM} + 4 c^2 v_\zeta^2}\,\right].
\label{eigenr1}
\end{eqnarray}
From $m^2_1$ we see that $\mu^2_{12}<0$, hence $m^2_1$ may have large mass. The SM-like Higgs boson may be identified with the scalar with mass $m_3$ in Eq.~(\ref{eigenr1}). In order to see this just make $v^2_\zeta = 0$.

The second mass matrix $M_{2R}$ in the basis  $(\textrm{Re}D^0_1,\textrm{Re}D^0_2)$ reads
\begin{equation}
\mathcal{M}^2_{2R}=\left(
\begin{array}{cc}
\mu^2_d+\frac{a^\prime}{2} v_{SM}^2+b_2v^2_\zeta & b_3v^2_\zeta\\
& \mu^2_d+\frac{a^\prime}{2} v_{SM}^2+b_2v^2_\zeta\\
\end{array}
\right) ,
\label{mr2}
\end{equation}
with the eigenvalues
\begin{equation}
\mathrm{m}^2_{R1}=2\mu^2_d+a^\prime v_{SM}^2+2(b_2-b_3)v^2_\zeta,\quad
\mathrm{m}^2_{R2}=2\mu^2_d+a^\prime v_{SM}^2+2(b_2+b_3)v^2_\zeta.
\label{eigenr2}
\end{equation}

The mass matrix in the $C\!P$ odd sector has the form in the basis $(\textrm{Im}D^0_1,\textrm{Im}D^0_2)$ (the would-be Goldstone bosons has been already decoupled)
\begin{equation}
M_{I}=\left(
\begin{array}{cc}
\mu^2_d+\frac{a^{\prime\prime}}{2} v_{SM}^2+b_2v^2_\zeta & b_3v^2_\zeta\\
& \mu^2_d+\frac{a^{\prime\prime}}{2} v_{SM}^2+b_2v^2_\zeta
\end{array}
\right),
\label{mi}
\end{equation}
with eigenvalues
\begin{equation}
m^2_{I1}=2\mu^2_d+a^{\prime\prime} v_{SM}^2+2(b_2-b_3)v^2_\zeta,\quad
m^2_{I2}=2\mu^2_d+a^{\prime\prime} v_{SM}^2+2(b_2+b_3)v^2_\zeta.
\label{eigeni1}
\end{equation}
Above we have defined $a^\prime=a_5+a_7+2a_6$ and $a^{\prime\prime}=a_5+a_7-2a_6$.

The model allows four neutral scalars with different masses that could contribute to the DM relic density in different proportion: two $C\!P$ even and two $C\!P$ odd.  Notice also that $a_6$ is the term in the scalar potential that transfer the $L$ violation to the active neutrino sector.

In the charged scalars sector, besides the charged would-be Goldstone boson, we have two charged scalar fields (we have already omitted the charged would-be Goldstone boson)
\begin{equation}
M^2_C=\left(\begin{array}{cc}
\mu^2_d+\frac{a_5}{2} v_{SM}^2+ b_2v^2_\zeta & b_3v^2_\zeta\\
& \mu^2_d+\frac{a_5}{2}v_{SM}^2+b_2v^2_\zeta
\end{array}\right),
\label{m4}
\end{equation}
with the non-zero eigenvalues given by
\begin{eqnarray}
m^2_{+1}=\mu^2_d+\frac{a_5}{2}v_{SM}^2+(b_2-b_3)v^2_\zeta \ , \
m^2_{+2}=\mu^2_d+\frac{a_5}{2}v_{SM}^2+(b_2+b_3)v^2_\zeta,
\label{mc}
\end{eqnarray}

Notice that
\begin{eqnarray}
m^2_{R1}-m^2_{+1}& = &\mu^2_d+\left(\frac{a_5}{2} + a_7+2a_6 \right)v^2_{SM} +(b_2-b_3)v^2_\zeta ,
\nonumber \\
m^2_{R2}-m^2_{+2}& = & \mu^2_d+\left(\frac{a_5}{2} + a_7+2a_6 \right)v^2_{SM} +(b_2+b_3)v^2_\zeta.
\label{dif1}
\end{eqnarray}
Notice that $\mu^2_d>0$ does not disappear in the mass difference above because, in oder to reproduce the Klein-Gordon equation for each component, a real scalar has a $1/2$ factor in the mass term related to the mass of a complex scalar.

\section{Lepton masses and the PMNS matrix}
\label{sec:masses}

\begin{figure}[!ht]
\subfloat[]{\includegraphics[width=5.5cm]{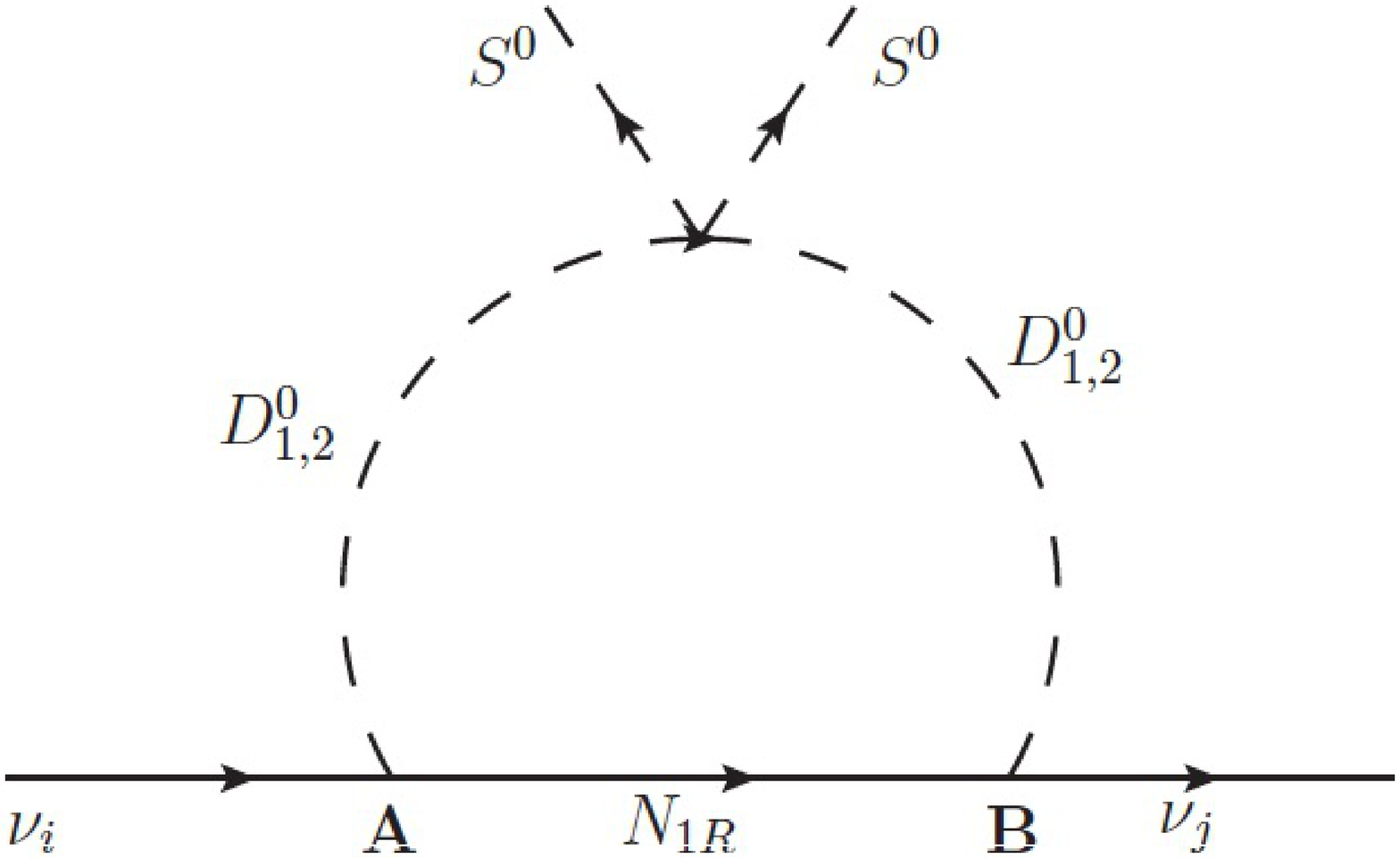}} \ \
\subfloat[]{\includegraphics[width=5.5cm]{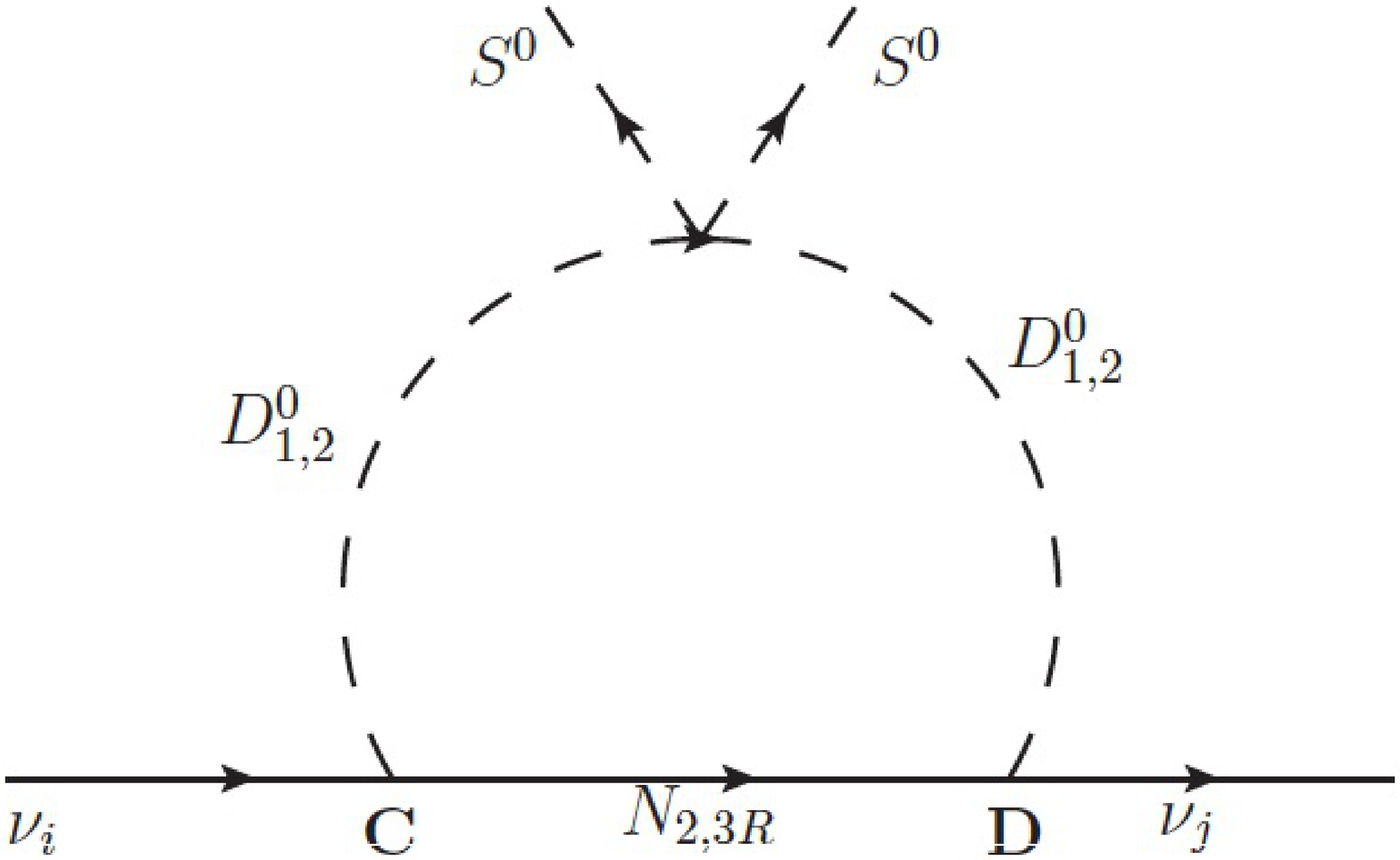}}  \ \
\caption{One-loop neutrino mass generation. Here $\textbf{A}$, $\textbf{B}$, $\textbf{C}$ and $\textbf{D}$ are the Yukawas given in Eq.~(\ref{yukawa1}), with $\textbf{A}=G_{id}^\nu$,
$\textbf{B}=G_{jd}^\nu$,  $\textbf{C}=v_\zeta G_{is}/\Lambda$, and $\textbf{D}=v_\zeta G_{js}/\Lambda$.}\label{loops}	
\end{figure}

In Sec.~\ref{sec:model} we have seen that at tree level the neutrinos are massless, the  $a_6$ term in Eq.~(\ref{potential3}) induce diagrams like those in Fig.~\ref{loops} and it is possible to implement the mechanism of Ref.~\cite{Ma:2006km} for radiative generation of neutrinos mass.
In fact, the diagram in Fig.~\ref{loops} are exactly calculable from the exchange of Re$D^0_{1,2}$ and
Im$ D^0_{1,2}$
\begin{equation}
(M_\nu)_{ij} =  \sum_{a,k} \frac{Y_{ik}Y_{jk} M_{k}}{32\pi^2} \left[ \frac{m^2_{Ra}}{m^2_{Ra}-M^2_k} \ln \frac{m^2_{Ra}}{M_{k}^2}- \frac{\mathrm{m}^2_{Ia}}{m^2_{Ia}-M^2_k} \ln \frac{m_{Ia}^2}{M_{k}^2}\right],
\label{numass1}
\end{equation}
where $a=1,2; k=s,d$; $m_{Ra}$ and $m_{Ia}$ are the masses of $\textrm{Re}D^0_{1,2}$ and $\textrm{Im}D^0_{1,2}$, respectively. In Eq.~(\ref{numass1}) $Y_{ik}Y_{jk}$ corresponds to $G^\nu_{is}G^\nu_{js}$ when the coupling is with the $N_s$; and  $Y_{ik}Y_{jk}$ corresponds to $G^\nu_{id}G^\nu_{jd}$ when the coupling is with the $N_d$, and finally $M_s$ is the mass of the right-handed neutrino $N_s$, and $M_d$ is the common mass of the  right-handed neutrino in the doublet of $S_3$, $N_d$ i.e, $M_2=M_3\equiv M_d$.
We can define $\Delta^2_a= m^2_{Ra}-~m^2_{Ia}=~4a_6v^2_{SM}$, and $m^2_{0a}=(m^2_{Ra}+m^2_{aI})/2$, $a=1,2$.

If $\Delta^2\ll m^2_{a0}$ we obtain
\begin{eqnarray}
(M_\nu)_{ij} &=& \frac{a_6v^2_{SM}}{16 \pi^2}\left[ \frac{G^\nu_{id}G^\nu_{jd} M_d}{m^2_{01} - M_d^2} \left(1 - \frac{M_d^2}{m^2_{01} - M_d^2} \ln \frac{m^2_{01}}{M_d^2} \right)+
\frac{G^\nu_{is} G^\nu_{js} M_s}{m ^2_{01}-M^2_s}
\left(1-\frac{M^2_s}{m ^2_{01}-M^2_s}\ln \frac{m^2_{01}}{M^2_s}\right)\right.\nonumber\\
&&\left. +
\frac{G^\nu_{id}G^\nu_{jd} M_d}{m^2_{02} - M_d^2} \left(1 - \frac{M_d^2}{m^2_{02} - M_d^2} \ln \frac{m^2_{02}}{M_d^2} \right)+
\frac{G^\nu_{is} G^\nu_{js} M_s}{m ^2_{02}-M^2_s}
\left(1-\frac{M^2_s}{m ^2_{02}-M^2_s}\ln \frac{m^2_{02}}{M^2_s}\right)
\right],
\label{numass2}
\end{eqnarray}
where $M_s$ is the mass of the right-handed neutrino $N_s$ and $M_d$ is the common mass of the neutrinos $N_d$. Under the condition in which the scalars are mass degenerated i.e.,  $b_3=0$ in (\ref{dif1}) we obtain just a factor 2 in Eq.(\ref{numass2}). Below, for simplicity, we will consider the case $b_3=0$.

In order to obtain the active neutrinos masses we assume a normal hierarchy and, without loss of generality, that $M_s \sim M_d$ and will be represented from now on by $M_R$. $M^\nu$ is diagonalized with a unitary matrix $V_L^\nu$ i.e.,  $\hat{M}^\nu~=~V^{\nu T}_LM^\nu V^\nu_L$,
where $\hat{M}^\nu = diag (m_1, m_2, m_3)\approx (0,\sqrt{\delta m^2_{12}}, \sqrt{\delta m^2_{23}})$. Taken the central values in PDG we have $\hat{M}^\nu\approx (0,x,y)$.

In the charged lepton sector we assume their masses at the central values in PDG  $\hat{M^l}  = (0.510, 105.658, 1776.86 )$ GeV. It is important to note from these considerations, that there exist a multitude of other possibilities which satisfy also the masses squared differences and the astrophysical limits in the active neutrino sector. Each one corresponds to different parameterization of the unitary matrices $V^l_{L,R},V^\nu_L$.

\begin{center}
\begin{figure}
\subfloat[]{\includegraphics[width=7.5cm]{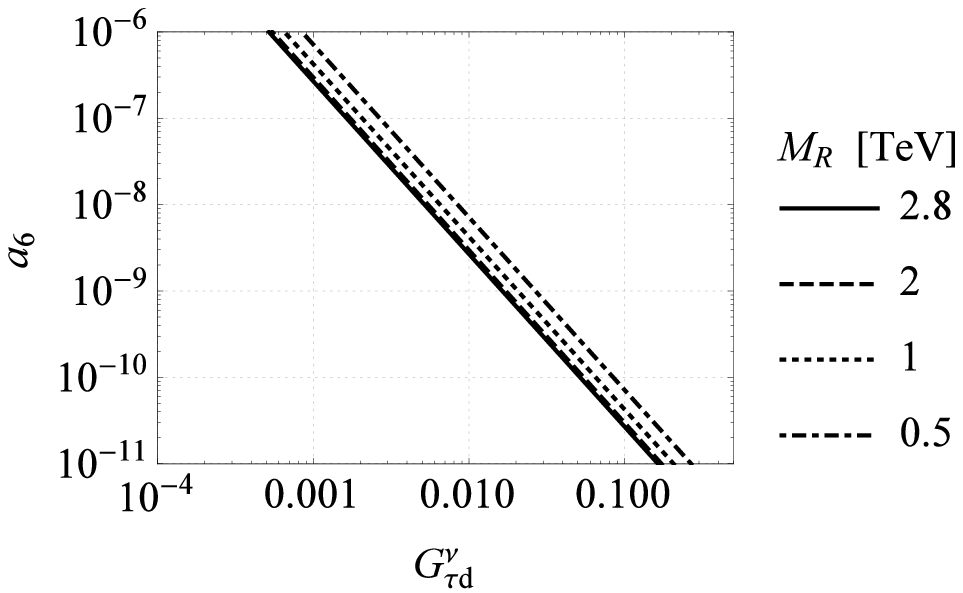}} \quad
\subfloat[]{\includegraphics[width=7.5cm]{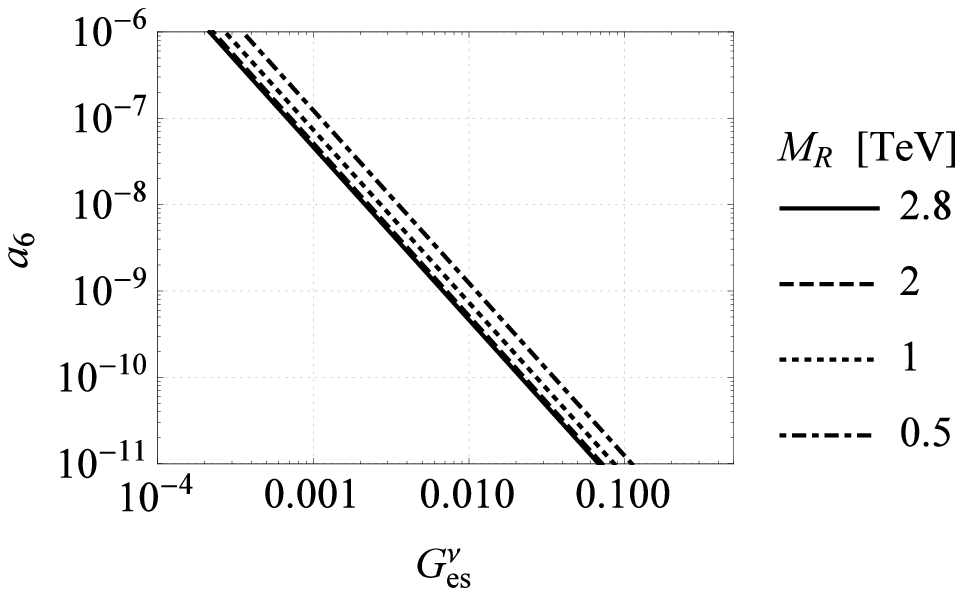}} \\
\subfloat[]{\includegraphics[width=7.5cm]{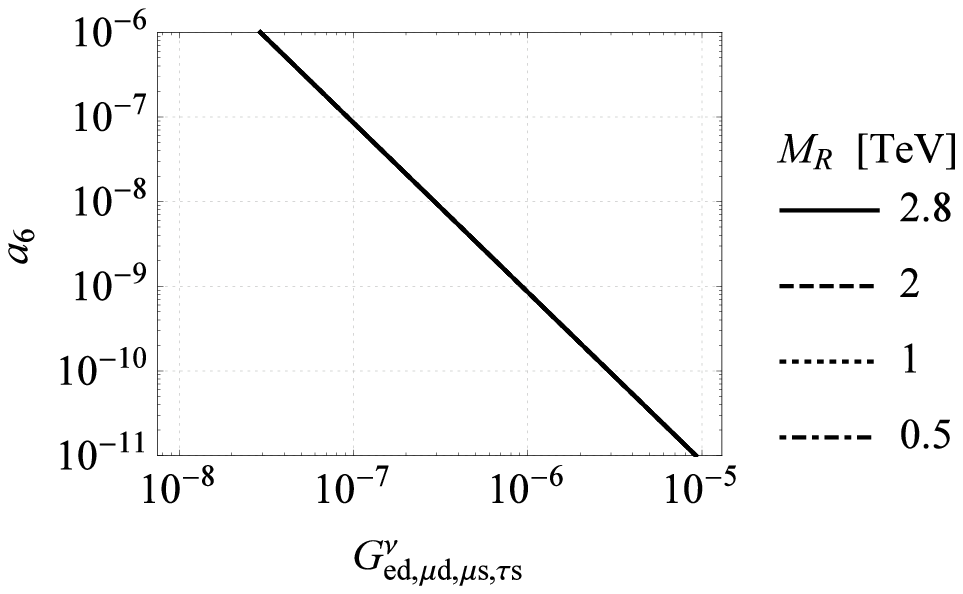}}
\caption{Dependence of $a_6$ with respect to the main Yukawas $G_{\tau d}^\nu$ (a) and $G_{es}^\nu$ (b) for fixed $M_R$ values. Notice that $G_{\tau d}^\nu\simeq G_{es}^\nu$. In (c) the Yukawas
$G_{ed,\mu d,\mu s,\tau s}^\nu$ provide the same value for any $M_R$ (overlapped curves).}
\label{FIGURE-Yukawas}
\end{figure}
\end{center}

We will obtain the neutrinos masses from Eq.~(\ref{numass2}). We have as free parameters $a_6$, $M_R$ and the Yukawas [$m_{+1}=m_{+2}\equiv m_+$ are also still free but they will enter only in the leptonic decays considered in Sec.~\ref{sec:decays}]. In the Fig.~\ref{FIGURE-Yukawas} we show the dependence of $a_6$ with respect to the main Yukawas $G_{\tau d}^\nu$ in (a) and $G_{es}^\nu$ in (b) for fixed $M_R$ values. Notice that $G_{\tau d}^\nu$ are essentially of the same order of magnitude, while the rest $G_{ed,\mu d,\mu s,\tau s}^\nu$  are suppressed by four orders of magnitude when comparing with any specific value of those $G_{\tau d,es}^\nu$.

\begin{table}[!h]
\begin{tabular}{|c|c|c|c|c|c|c|c|}\hline
$V_L^l$ parameterization &
Masses in TeV & $G^l_{11} $  & $G^l_{12}$ & $G^l_{13}$  & $G^l_{22}$ & $G^l_{23} $ &  $G^l_{33}$  \\ \hline
P1 & $M_R=2.8 $, $M_0 = 2.2 $
& 0.000421836  &  0.000514741 & -0.000800772 &   0.00374767 & -0.00356758 & 0.00367645   \\ \hline
P2 & $M_R=2\ \ $, $M_0 = 2.2 $
& 0.000421864  &  0.000514825 & -0.000800852 &   0.00374768 & -0.00356756 &  0.00367641  \\ \hline
P3 & $M_R=1.5 $, $M_0 = 2.2 $
& 0.000421896  &  0.000514922 & 0.000800945 &   0.0037477 & -0.00356754 &  0.00367636  \\ \hline
P4 & $M_R=1\ \ $, $M_0 = 2.2 $
& 0.00042195  &  0.000515086 & -0.000801103 &   0.00374772 & -0.00356751 &  0.00367628  \\ \hline
P5 & $M_R =  0.5 $, $M_0 = 2.2 $
& 0.00042205  &  0.000515393 & -0.000801398 &   0.00374778 & -0.00356744 &  0.00367612  \\ \hline
\end{tabular}
\caption{Masses of the scalars in Scotogenic model (in GeV).}
\label{table4}
\end{table}

The mass matrices in the charged lepton sector $M^l$ are diagonalized by a bi-unitary transformation $\hat{M}^l=V^{l\dagger}_L M^l V^l_R$ and $\hat{M}^l = diag (m_e, m_\mu, m_\tau)$. The relation between symmetry eigenstates (primed) and mass (unprimed) fields are $l^\prime_{L,R}=V^l_{L,R}l_{L,R}$
and $\nu^\prime_L=V^\nu_L \nu_L$, where
$l^\prime_{L,R}=(e^\prime,\mu^\prime,\tau^\prime)^T_{L,R}$, $l_{L,R}=(e,\mu,\tau)^T_{L,R}$,  $\nu ^\prime_L=(\nu_e,\nu_\mu,\nu_\tau)^T_L$
and $\nu_L=(\nu_1,\nu_2,\nu_3)_L$.
Defining the lepton mixing matrix as $V_{PMNS}=V^{l\dagger}_LV^\nu_L$, it means that this matrix appears in the charged currents coupled to $W_\mu^+$. We have tested the robustness of our fitting of the lepton masses and the leptonic mixing matrix by using several parametrization corresponding to the values of the Yukawa couplings given in Table~\ref{table4}. We omit the respective matrices $V^l_L$ and $V^\nu_L$ but in all cases we have obtained:
\begin{equation}
\vert V_{PMNS}\vert \approx\left(\begin{array}{ccc}
0.815 & 0.565 & 0.132\\
0.479 & 0.527 & 0.702\\
0.327 & 0.635 & 0.700\\
\end{array}\right),
\label{pmns}
\end{equation}
which is in agreement within the experimental error data at 3$\sigma$ given by~\cite{GonzalezGarcia:2012sz}
\begin{equation}
\vert V_{PMNS}\vert \approx\left(\begin{array}{ccc}
0.795-0.846& 0.513-0.585 & 0.126-0.178\\
0.4205-0.543 & 0.416-0.730  & 0.579 - 0.808 \\
0.215 - 0.548 & 0.409 - 0.725 & 0.567 -0.800 \\
\end{array}\right),
\label{pmnsexp}
\end{equation}
and we see that it is possible to accommodate all lepton masses and the PMNS matrix. Here we do not
consider $CP$ violation.


\section{Dark Matter}
\label{sec:dm}

As we said before, the present model may have a multi-component DM spectrum, which means that many particles may contribute to the relic density of DM, but we will consider the simplest example where one of the $C\!P$ even scalar, say $R_1$, and one of the $C\!P$ odd scalar, say $I_1$, as the dark matter candidates, each case is considered separately for simplicity. A two inert doublet model without right-handed neutrinos and scalar singlet was considered in Ref.~\cite{Fortes:2014dca}.

As usual, in order to determine the relic density, we solve the Boltzmann equation. Firstly, considering $R$ as the candidate, we have
\begin{equation}\label{dm1}
\frac{dn_R}{dt}+3Hn_{R}=-\langle \sigma |v| \rangle [(n_R)^{2}-(n_{R}^{eq})^{2}],
\end{equation}
where $\langle \sigma |v| \rangle$ is the annihilation cross section already thermally averaged and $H$ is the Hubble constant.
In the thermal equilibrium, the number density of DM \cite{Beltran:2008xg} is
\begin{equation}\label{dm2}
n^{eq}_R=g\left(\frac{m_R\,T}{2\pi}\right)^{3/2}\exp\left(-\frac{m_{R}}{T}\right),
\end{equation}
where $g=1$ for a scalar DM.
When solving the Boltzmann equation we obtain the equation for the relic density:
\begin{equation}\label{dm3}
\Omega_{R} h^{2}\approx \frac{1.04\times 10^{9}x_{F}}{M_{Pl}\sqrt{g_{*}}(a+3b/x_{F})},
\end{equation}
where $M_{Pl}=1.22\times 10^{19}$ GeV is the Planck mass, $x_{F}= m_{R}/T_{F}$, where $m_R$ is the mass of the neutral scalar and $T_{F}$ is the temperature at freeze-out, the terms $a$ and $b$ result from the partial wave expansion of $\sigma |v|= a + bv^{2}$. The number of relativistic degrees of freedom $g_{*}=118.375$  is a result of the SM particles plus three right-handed neutrinos, five neutral scalars, two pseudo-scalars and two charged scalars.
The evaluation of $x_{F}$ leads to
\begin{equation}
\label{dm4}
x_{F}=\ln \left [\widetilde{c}(\widetilde{c}+2)\sqrt{\frac{45}{8}}\frac{g m_{R}M_{Pl}(a+6b/x_{F})}{2\pi^{3}\sqrt{g_{*}(x_{F})}}  \right ],
\end{equation}
where the  unitary parameter $\widetilde{c}\approx 5/4$.

Here we will consider the solution for the relic density which, at the same time, solves the charged lepton masses and neutrino masses given in Eq.~(\ref{numass1}) for the sets of parameters showed in Tables \ref{table2} and \ref{table3}, in order to obtain the PMNS matrix.  We call them scenario 1 and 2, when $R_1$ and $I_1$ is the DM candidate, respectively. In both scenarios the Yukawas values adjust the squared masses differences for the neutrinos and the PMNS.

\begin{table}[!h]
\begin{tabular}{|c|c|c|c|c|c|c|}\hline
Scenario & $G^\nu_{es}$ & $G^\nu_{\mu s}$ & $G^\nu_{\tau s}$ & $G^\nu_{ed}$ & $G^\nu_{\mu d}$ &  $G^\nu_{\tau d}$ \\ \hline
 1
& 1.17$\times 10^{-7}$  & $10^{-11}$ & $10^{-11}$ &  $10^{-11}$ & $10^{-11}$ & 2.81$\times 10^{-7}$   \\ \hline
 2
& 1.56$\times 10^{-7}$  &  $10^{-11}$ & $10^{-11}$ &   $10^{-11}$ & $10^{-11}$ & 3.75$\times 10^{-7}$  \\ \hline
\end{tabular}
\caption{The Yukawas values for Eq.~(\ref{numass2}) that solve DM.}
\label{table2}
\end{table}

\begin{table}[!h]
	\begin{tabular}{|c|c|c|c|c|c|c|c|c|c|c|c|}\hline
Scenario		& $M_{+1}$  & $M_{+2}$ & $m_{\zeta1}$ & $m_{\zeta2}$ & $M_{R1}$ &  $M_{R2}$ & $M_{I1}$& $M_{I2}$ & $M_d$ & $M_s$ & $a_6$  \\ \hline
1  & 109.02  &  1477.64 & 749.81 &  3747.26 & 85.20 & 2085.74   & 161.84 & 2090.27 &  240  & 240 & -1.16$\times 10^{-1}$\\ \hline
2 & 109.12 & 1477.65 & 749.81  & 3747.26 & 257.25 & 2099.82  & 113.70 & 2087.10 &  240 & 240 & 2.6$\times 10^{-1}$ \\ \hline
	\end{tabular}
\caption{Masses (in GeV) of the scalars in this model. $a_6$ is a dimensionless coupling in the non-Hermitian quartic scalar interaction
		which transfer the $L$ violation to the active neutrino sector.}
	\label{table3}
\end{table}

With the numbers in Tables~\ref{table2} and \ref{table3} we obtain the mixing matrix, that is:
\begin{equation}
\label{numixingL}
V^{\nu}_L \approx  \left(\begin{array}{ccc} 0.00010637 & -1 & 0.0000444397 \\  -1 & -0.000106374857 & 0.00007634160462 \\ 0.000076336876851 & 0.000044447836932 & 1 \end{array}\right),
\end{equation}
and the charged leptons has the following Yukawas: $G^l_{11}=0.0004219$, $G^l_{12} = 0.000515$,
$G^l_{22} = 0.00374772$,  $G^l_{13} = -0.000801115$, $G^l_{23} = -0.0035675$, $G^l_{33} = 0.00367627$,
and we obtain $m_e = 0.510$ MeV, $m_\mu = 105.658$ MeV and $m_\tau = 1776.86 $ MeV, and the mixing matrix is
\begin{equation}
\label{lmixingL}
V^{l}_L =  \left(\begin{array}{ccc} 0.564902 & 0.52704 & 0.634914  \\ 0.814515 & -0.479341& -0.326799 \\ -0.132104 & -0.701756 & 0.700062 \end{array}\right).
\end{equation}
As previously defined the mixing matrix for the leptonic sector $V_{PMNS}=V^{l\dagger}_LV^\nu_L$, From Eqs.~(\ref{numixingL}) and (\ref{lmixingL}) we obtain again Eq.(\ref{pmns}).

To perform DM calculation we have used \texttt{MicrOmegas} package \cite{micromegas}. For instance, let us consider scenario 1, where $R_1$  is the DM candidate. In the range of parameters used by us, DM annihilates mainly in $W^{+}W^{-}$.  Once again we emphasize that other solutions in other annihilation channels do exist. We have chosen the following parameters for the couplings, and vacuum expected value: $G_{\tau d} = 2.81\times10^{-7}$, $G_{\tau s} = 1 \times 10^{-11}$, $v_\zeta\lesssim \Lambda = 1000$ GeV, $\lambda^{\prime}=2.7\times 10^{-2}$, $\lambda^{\prime\prime}=0.34$, $\mu^2_{d}=2.809\, (\textrm{TeV})^2$, for values of other parameters see Table~\ref{table2}. With this parameters choice $m_{R1} = 85.15$ GeV. The  parameter dependence of $m_{R1}$ is presented in the last sections. So, the dominant contributions for $\Omega$ are 99\% in $R_{1}R_{1}\rightarrow W^{+}W^{-}$. In this case, $x_{F}\sim 23.8$. The annihilation cross section is $\langle\sigma v\rangle= 1.0\times 10^{-26}$ cm$^{3}/$s and the DM-nucleus cross section for spin-independent elastic scattering is numerically given by $\sigma_{SI}^{p}=5.84\times 10^{-46}$ cm$^{2}$ and $\sigma_{SI}^{n}=6.70\times 10^{-46}$ cm$^{2}$.

In scenario 2 we consider $I_{1}$ as the DM candidate. In this case, as a result of the parameter choice ($G_{\tau d} =  3.75\times10^{-7}$, $G_{\tau s} = 1 \times 10^{-11}$, $v_\zeta\lesssim\Lambda = 1000$ GeV, $\lambda^{\prime}=1$, $\lambda^{\prime\prime}=0.12$, $\mu^2_{d}=2830.24$ TeV$^2$),  $I_{1}$ annihilates 97\% in $I_{1} I_{1}\rightarrow hh$ and 2\% in $I_{1} I_{1}\rightarrow b\overline{b}$. The value of other parameters can be seen in Table~\ref{table2}.
The annihilation cross section is $\langle\sigma v\rangle = 4.55\times 10^{-28}$ cm$^{3}/$s and the DM-nucleus cross section for spin-independent elastic scattering is numerically given by $\sigma_{SI}^{p}=6.40\times 10^{-45}$ cm$^{2}$ and $\sigma_{SI}^{n}=7.35\times 10^{-45}$ cm$^{2}$.

In these two scenarios, we had set $R_{1}$ and  $I_{1}$ as DM candidates, making them lighter than the others possible neutral scalars. We emphasize that other choices for DM are possible so that other annihilation channels may also give interesting signatures. There is also the possibility that two, three or even four of the neutral scalars contribute partially to the DM density, but this case is beyond the scope of this paper.

\begin{figure}[ht]
\subfloat[]{\includegraphics[width=8cm]{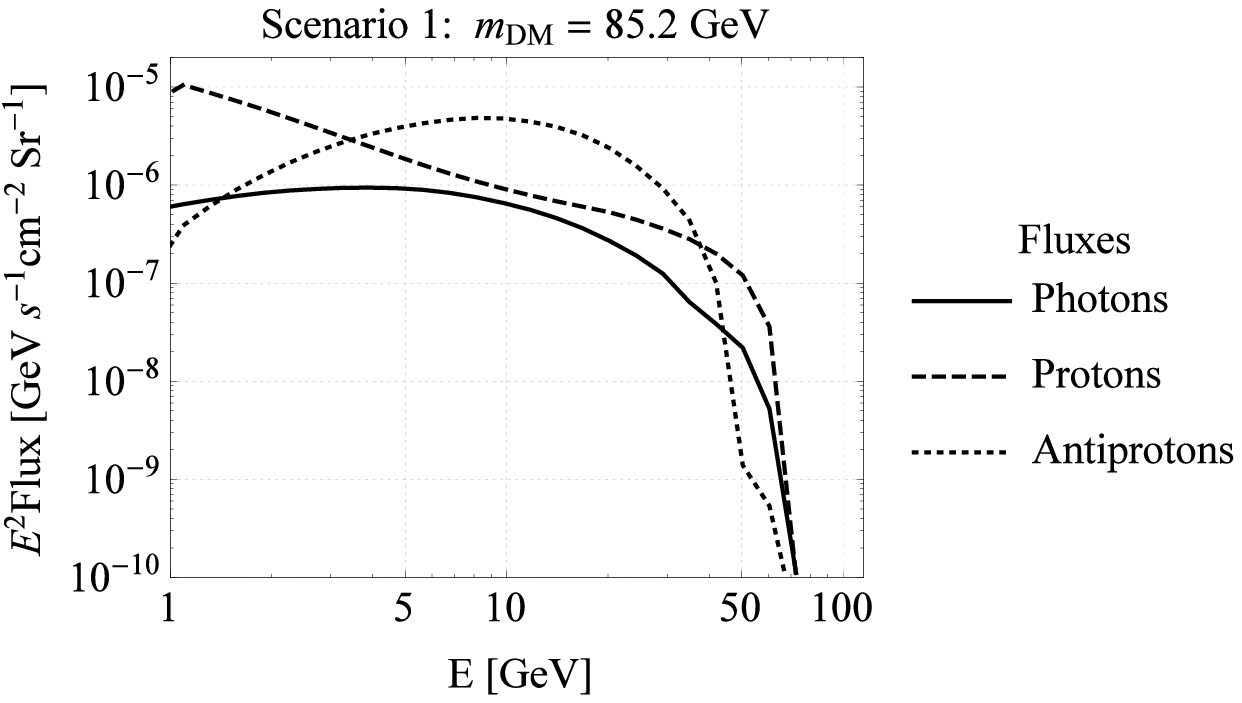}} \quad
\subfloat[]{\includegraphics[width=8cm]{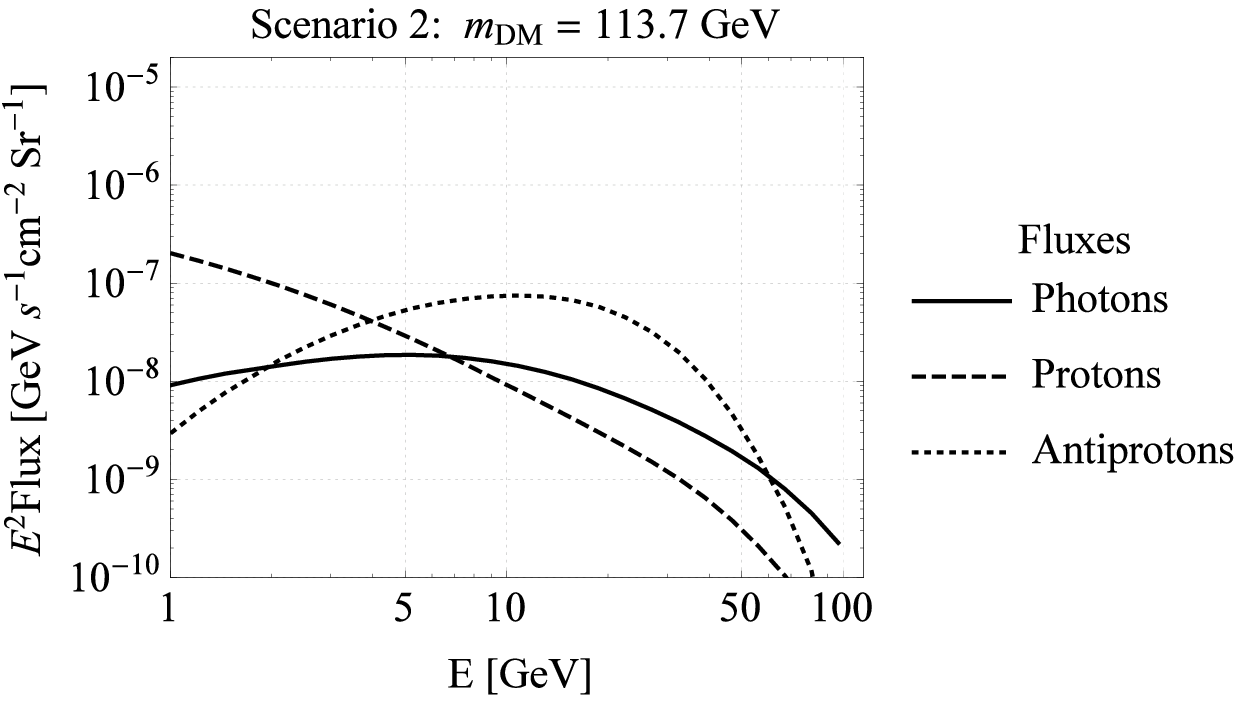}}
\caption{Prediction of fluxes for photons, protons, and anti-protons in the scotogenic model. Scenarios (a) 1 and (b) 2.}\label{FIGURE-fluxes}
\end{figure}

In Fig.~\ref{FIGURE-fluxes} we present the fluxes of photons, positrons and antiprotons in the scenario 1 with $m_{DM}=85.15$ GeV and the scenario 2 with $m_{DM}=113.70$ GeV, where the upper limits of the energy spectrum are determined by the DM masses since annihilation occurs near at rest.
The model can accommodate DM candidtes with smaller masses than the values above.

\section{The leptonic decays $l_i\to l_j\gamma$ and $l_i\to l_jl_k\bar{l}_k$.}
\label{sec:decays}

In this section we study the impact of the new particles, the charged scalars
$D_{1,2}^+$ and the right-handed neutrinos $N_{s,2,3}$ in the lepton flavor violating processes $l_i\to l_j\gamma$.
Here we will consider these rare decays in two cases: one in which we do not care with DM solutions and one in which we use the parameters for having a DM candidates that also give the correct lepton masses and the PMNS.

In terms of the leptons mass eigenstates, the interactions with charged scalars from Eq.~(\ref{yukawa1}) are written as
\begin{eqnarray}\label{yukawa2}
\mathcal{L}^{l-N}_{Yukawa}&=&-\bar{l}_{kL}V^{l\dagger}_{ki}\left[ G^\nu_{id}(N_{2R}D^-_1+N_{3R}D^-_2)+\frac{v_\zeta}{\Lambda} G^\nu_{is} N_s(D^-_1-D^-_2)\right],
\end{eqnarray}
$i,k=e,\mu,\tau$,
the values of the entries of the matrices $G_{i,d}^\nu$ and $G_{i,s}^\nu$ are given in Table~\ref{table2}.

In the model, the allowed lepton flavor violation (LFV) decays $l_i\to l_j\gamma$ and $l_i\to l_jl_k\bar{l}_k$  arise only at the  1-loop level. These diagrams  are generated by the known SM contribution $W~\&~\nu_l$ and by the new content $D_1^+ \& N_2$, $D_2^+ \& N_3$, $D_1^+\& N_s$ and $D_2^+\& N_s$.

For $l_i\to l_j\gamma$, it is known that the SM contribution is extraordinarily suppressed with respect to the experimental capabilities of detection, see Table~\ref{Table-loop-experiments}. As we will show below, the new particle content in the model predicts signals close to the experimental upper limits for the space of our considered allowed parameters. Regarding the three body decay $l_i\to l_jl_k\bar{l}_k$, it arises when in $l_i\to l_j\gamma$ we attach to the photon the $\gamma l\bar{l}$ coupling. In the following we are interested in presenting the $\mu\to ee\bar{e}$ channel, because it provides interesting results near the experimental upper limit, while all the other channels are out of the experimental interest region because
the devices are unable of reaching such suppressed signals.

In our study we have solved the amplitudes and the loop integrals with the help of \texttt{Mathematica}, \texttt{FeynCalc}~ \cite{Mertig:1990an,Shtabovenko:2016sxi}, and \texttt{Package-X}~\cite{Patel:2015tea}.

\subsection{Predictions of $\mu\to e\gamma$ and $\mu\to ee\bar{e}$  in the scotogenic model without dark matter.}
\label{subsec:muegamma}

We start our numerical analysis of the decays in the scotogenic model without DM content. Accordingly to the Yukawa values derived in the Sec.~\ref{sec:masses} (see Fig.~\ref{FIGURE-Yukawas}), the obtained values of the Yukawas are $G_{ad,as}^\nu\in [10^{-11},10^{-1}]$, they satisfy the neutrino masses.
Notice that $G_{\tau d}^\nu\simeq G_{es}^\nu \gg G_{ed,\mu d,\mu s,\tau s}^\nu$.
We recall that all the previous analyse were done in the case of $b_3=0$ in which the charged scalars are mass degenerated. As starting point we consider the mass of the charged scalar in the range $m_{+}\in$ [80,750] GeV, $N_{2,3,s}$ degenerated as well with values $m_N\in$ [250, 4000] GeV.
We have tested the five different parameterizations of the $V_L^l$ matrices derived in the Sec.~\ref{sec:masses} (see Table~\ref{table4}). They can be separated into two sets which will have two different behaviours in the processes, the set A is conformed by the parameterizations P1 and P5, and the B by P2, P3 and P4.

\begin{center}
\begin{figure}[h]
\subfloat[]{\includegraphics[width=8.5cm]{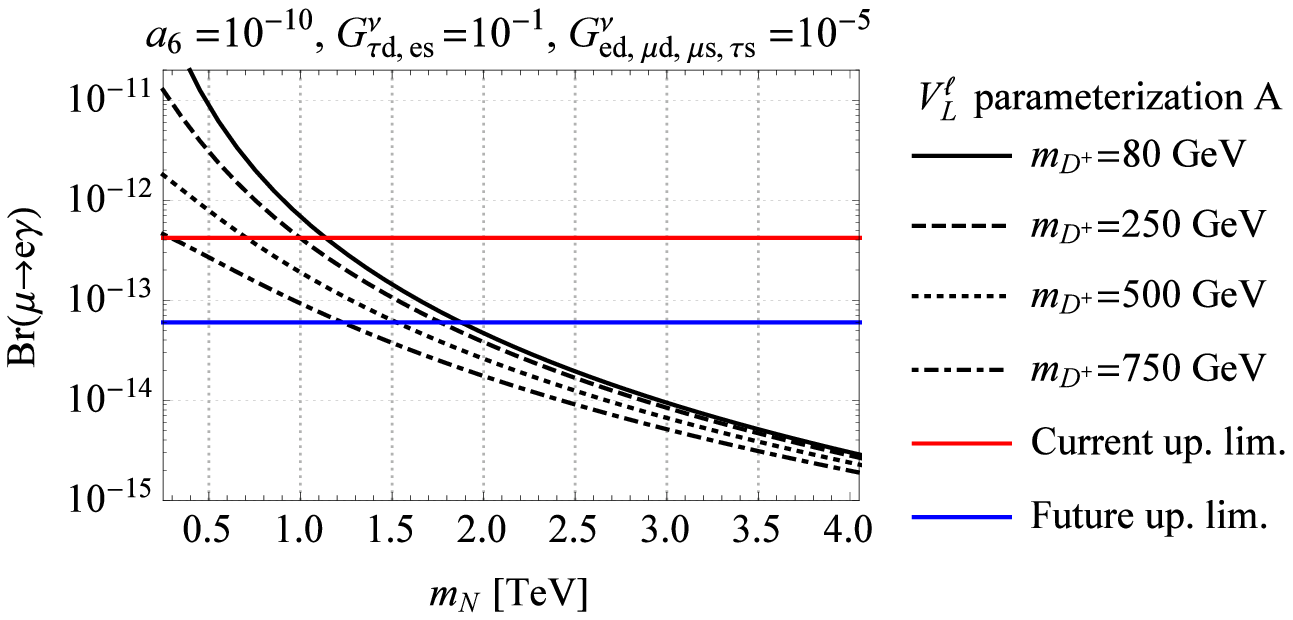}} \quad
\subfloat[]{\includegraphics[width=8.5cm]{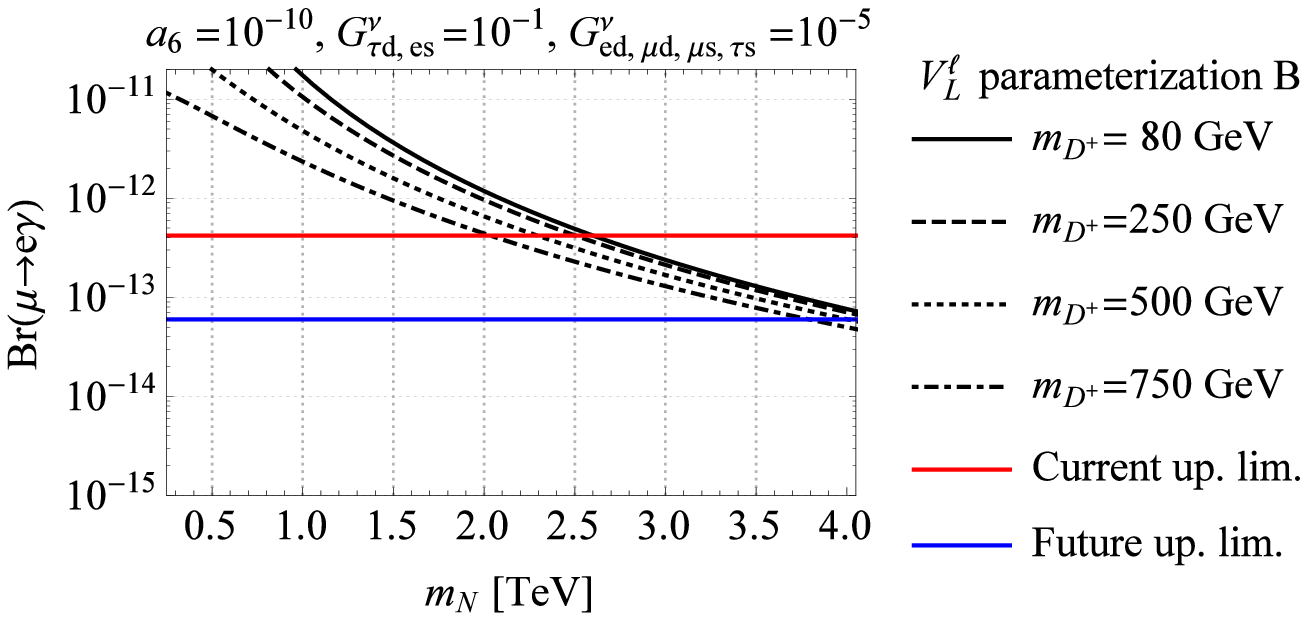}} \\
\subfloat[]{\includegraphics[width=8.5cm]{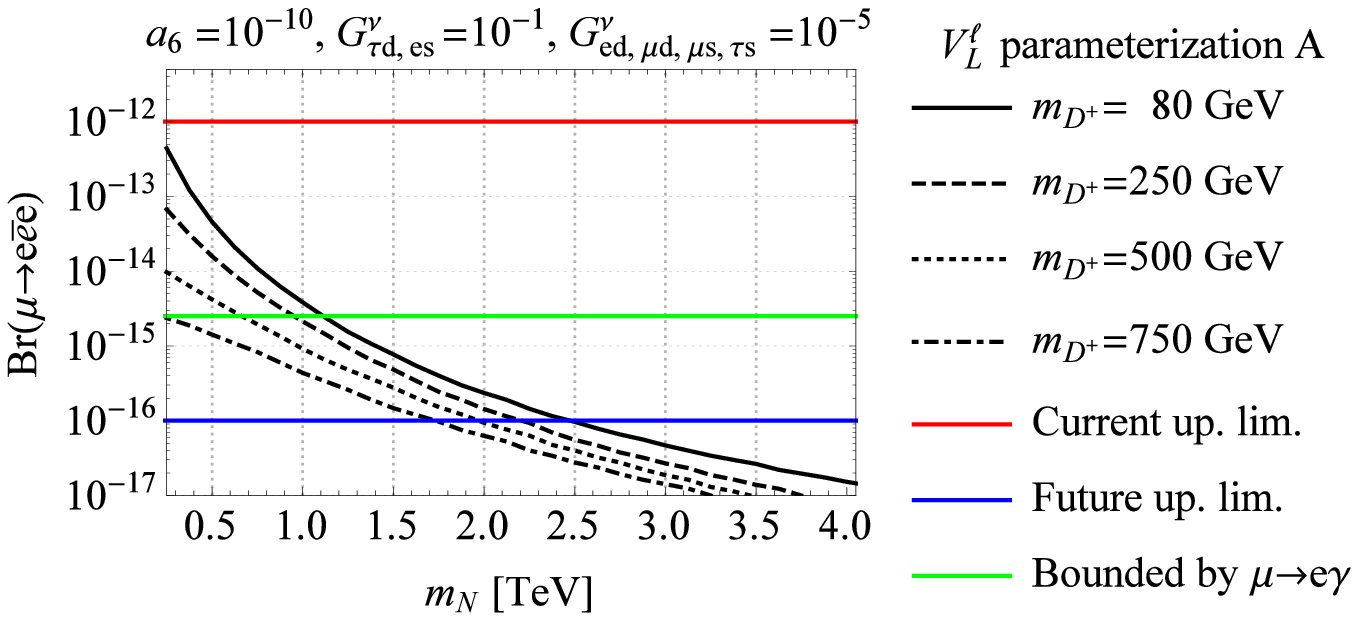}} \
\subfloat[]{\includegraphics[width=8.5cm]{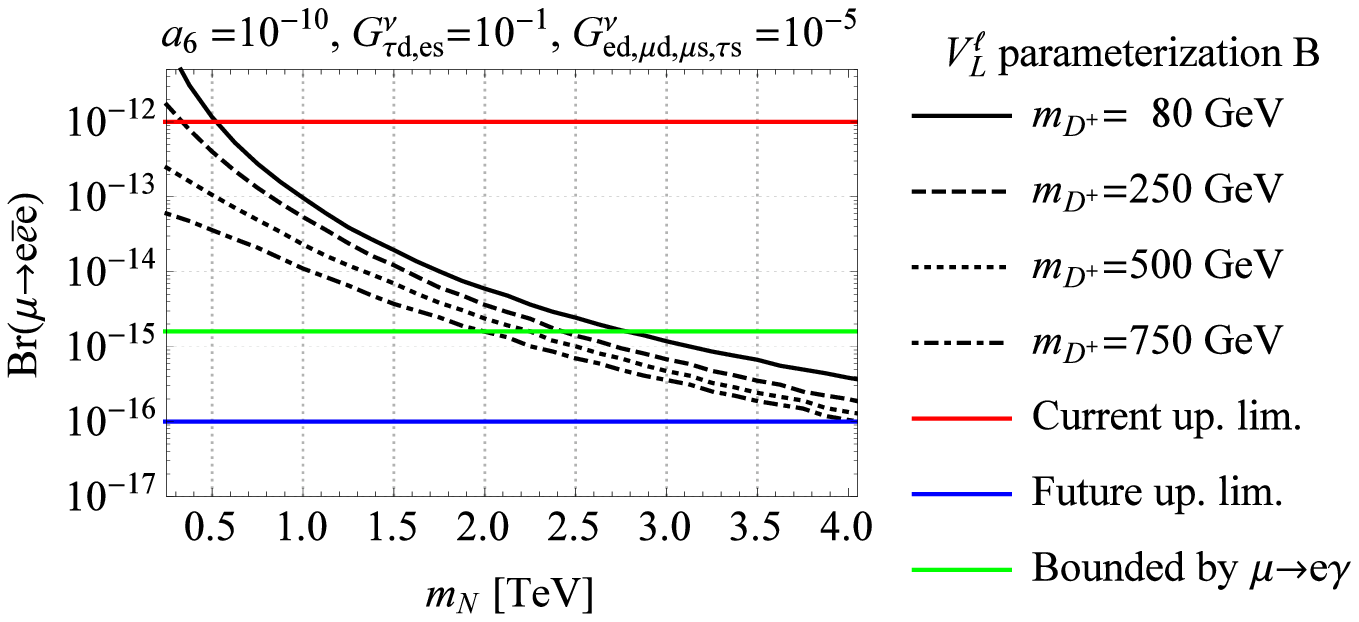}}
\caption{Decays $\mu\to e\gamma$ in (a)-(b) and $\mu\to ee\bar{e}$ in (c)-(d) in the scotogenic model without DM, with $G_{\tau d,es}^\nu=10^{-1}$ and $G_{ed,\mu d,\mu s,\tau s}^\nu=10^{-5}$.}
\label{FIGURE-decays-G01}
\end{figure}
\end{center}

For the channel $\mu\to e\gamma$ such situation occurs when $G_{\tau d}^\nu= G_{es}^\nu\sim10^{-1}$ and $G_{ed,\mu d,\mu s,\tau s}^\nu\sim10^{-5}$. In the Fig.~\ref{FIGURE-decays-G01}~(a)-(b) it is shown the
Br$(\mu\to e\gamma)$ as function of the sterile neutrino mass $m_N\in[250,4000]$ GeV with given values for the charged scalar $m_{D^+}=80, 250, 500, 750$ GeV.
The current experimental upper limit Br$(\mu\to e\gamma)^\text{Exp}<4.2\times10^{-13}$, indicated with the red line in the plots, will constrain the right-handed neutrino mass $m_N$ for given values of $m_{+}$ in order to respect such limit, those constraints are listed in the Table~\ref{TABLE-mass-bounds},
where it is evident that the $V_L^l$ parameterization A in Fig.~\ref{FIGURE-decays-G01}~(a) allows a lighter mass for the right-handed neutrino than the parameterization B in Fig.~\ref{FIGURE-decays-G01}~(b).
\begin{table}[ht]
  \centering
\begin{tabular}{|c|c|c|c|c|}\hline
Decay                  & Current limit         & Future limit         & SM \\
\hline
Br($\mu\to e\gamma$)   & $< 4.2\times10^{-13}$ \cite{TheMEG:2016wtm} & $<6.0\times10^{-14}$ \cite{Baldini:2013ke} & $10^{-48}$ \\
Br($\tau\to e\gamma$)  & $< 3.3\times10^{-8\ }$ \cite{Olive:2016xmw} & $<3.3\times10^{-9\ }$ \cite{Aushev:2010bq} & $10^{-49}$ \\
Br($\tau\to\mu\gamma$) & $< 4.4\times10^{-8\ }$ \cite{Olive:2016xmw} & $<3.3\times10^{-9\ }$ \cite{Aushev:2010bq} & $10^{-49}$ \\
\hline
\end{tabular}
\caption{$l_i\to l_j\gamma$, experimental upper limits and the SM predictions.}\label{Table-loop-experiments}
\end{table}

\begin{table}[ht]
\centering
\begin{tabular}{|c|c|c|}
\hline
\multicolumn{3}{|c|}{$\mu\to e\gamma$} \\
\cline{1-3}
\multirow{2}{*}{$m_{D^+}$ [GeV]} & \multicolumn{2}{c|}{$m_N$ [GeV]} \\
\cline{2-3}
 & $V_L^l$ parameterization A & $V_L^l$ parameterization B  \\
\hline
80  & $>$1140 & $>$2610  \\
250 & $>$1000 & $>$2510  \\
500 & $>$~710 & $>$2300  \\
750 & $>$~295 & $>$2040  \\
\hline
\end{tabular}
\caption{Mass constraints for the right-handed neutrino with fixed values of $m_{D^+}$ in order to respect
Br$(\mu\to e\gamma)^\text{Exp}<4.2\times10^{-13}$, here $G_{\tau d}^\nu= G_{es}^\nu\sim10^{-1}$ and $G_{ed,\mu d,\mu s,\tau s}^\nu\sim10^{-5}$.}\label{TABLE-mass-bounds}
\end{table}

\begin{table}[ht]
\centering
\begin{tabular}{|c|c|}
\hline
\multicolumn{2}{|c|}{Br$(\mu\to ee\bar{e})$} \\
\cline{1-2}
 $V_L^l$ parameterization A & $V_L^l$ parameterization B  \\
\hline
$<2.2\times10^{-15}$ & $<1.4\times10^{-15}$ \\
\hline
\end{tabular}
\caption{Br$(\mu\to ee\bar{e})$ predictions from the Br$(\mu\to e\gamma)$ constraints.}\label{TABLE-mu-3e-bounds}
\end{table}

Regarding to the subcase $\mu\to ee\bar{e}$, we are able to predict the branching ratio
from the neutrino right-handed mass constraints obtained for the $\mu\to e\gamma$ channel. These predictions are organized in the Table~\ref{TABLE-mu-3e-bounds}, and such values are indicated with the green line in the Fig.~\ref{FIGURE-decays-G01}~(c) and (d), being of the same order of magnitude $\sim10^{-15}$ for both scenarios.

About the analogous tau decays, for the same space of parameter values than in the $\mu\to e\gamma$ case, and respecting the obtained mass bounds, we have found that Br$(\tau\to e\gamma)\leq10^{-12}$ and Br$(\tau\to e\gamma)\leq10^{-14}$, which are beyond the current and upcoming experimental capabilities of detection, see Table~\ref{Table-loop-experiments} for comparison.

\subsection{Predictions of $l_i\to l_j\gamma$ in the scotogenic model with dark matter}
\label{subsec:meee}

As commented in the Sec.~\ref{sec:dm}, the model can be extended to include DM. In order to estimate the consequences on the transition $\mu\to e\gamma$, we consider the Yukawa values given in Table~\ref{table2}. The resulting prediction with $G_{\tau d}^\nu=G_{es}^\nu=10^{-7}$ is
\begin{equation}\label{}
\text{Br}(\mu\to e\gamma)=10^{-34},
\end{equation}
which is beyond the scope of detection.

\section{Conclusions}
\label{sec:con}

Here we have considered an extension of the SM with three scalar doublets of $SU(2)$ with $S_3$ and $\mathbb{Z}_2$ symmetries. We had analysed all the mass spectra in the scalar sectors and used the scotogenic mechanism for generating neutrino masses. Moreover, we had obtained the PMNS matrix once the unitary matrices which diagonalize the lepton masses are obtained. Although the model can have many DM candidates, we have shown two cases in which the DM candidate is a $C\!P$ even scalar (scenario-1) and other one in which the DM is composite of $C\!P$ odd scalar (scenario-2).  But we emphasize that other possible choices  for DM candidates are possible, considering for example, smaller masses, since besides the SM-like scalar, we have eight additional neutral scalars in the model. The study of other candidates and other channels of annihilation will be done soon. We had exemplified in some range of parameters space, two DM candidates for the model. For the scenarios 1 and 2 presented, DM annihilates mainly in $W^{+}W^{-}$ and $hh$ respectively. We have also presented some fluxes for this model. Of course, there may be other possible scenarios which could explain the Galactic gamma ray excess, as well as the the PAMELA and AMS-02 results. These processes could tightly constrain the parameter space of this sort of scotogenic models.

The considered scotogenic model without DM provides optimistic predictions for possible detection of the LFV decay $\mu\to e\gamma$ due to our solution space of the Yukawa values, which adjusts the squared masses differences for the neutrinos and the PMNS matrix. Our estimations predict a mass for the right-handed neutrino  starting from $m_N>295$ GeV, and from $\mu\to e\gamma$ we predict Br$(\mu\to ee\bar{e})\lesssim 10^{-15}$.
On the other hand, considering DM content in the model we found that Br$(\mu\to e\gamma)\sim 10^{-34}$, which is out of detection range.

\acknowledgments

ACBM thanks CAPES for financial support, JM thanks to FAPESP for financial support under the processe
number 2013/09173-5, and  VP thanks to CNPq  for partial financial support.
\appendix

\newpage


\begin{figure}[!h]
\subfloat[]{\includegraphics[width=5.5cm]{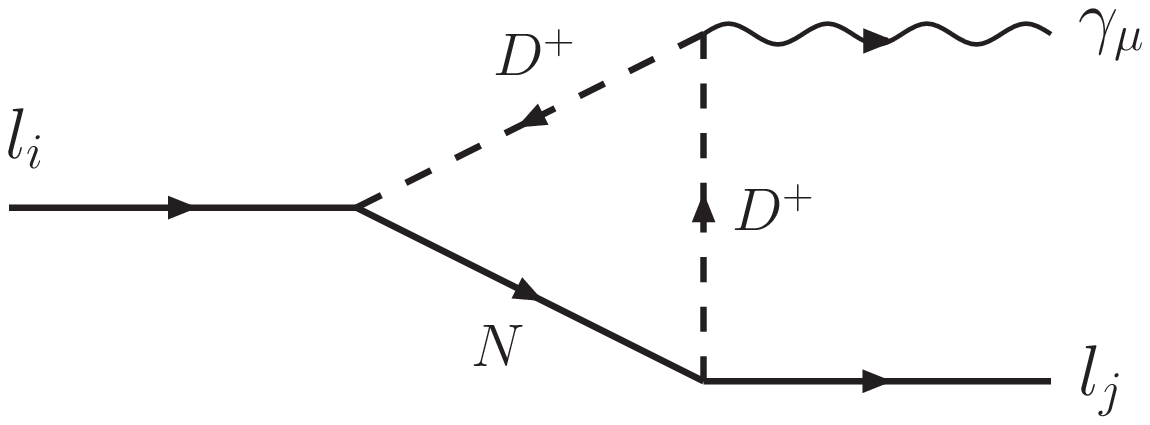}} \qquad
\subfloat[]{\includegraphics[width=5.5cm]{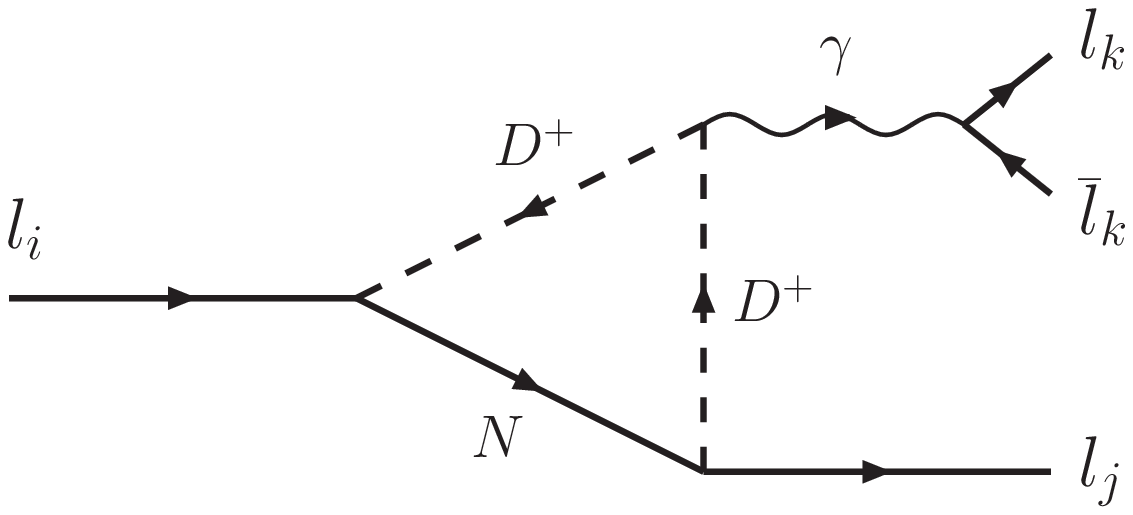}}
\caption{Decays (a) $l_i\to l_j\gamma$ and (b) $l_i\to l_jl_k\bar{l}_k$ in the scotogenic model. Generic sample contributions from a sterile neutrino $N$ and a charged scalar $D^+$.}
\label{FIGURE-loops}
\end{figure}

\newpage

\end{document}